\documentclass[doublecol]{epl2}

\usepackage{amsmath}

\title{Time Scales in the Theory of Elasto-Plasticity of Amorphous Solids}
\shorttitle{Title} 

\author{Laurent Bou\'e\inst{1}, Peter Harrowell\inst{2}, Smarajit Karmakar\inst{1}, Edan Lerner\inst{1},
Itamar Procaccia\inst{1}, Ido Regev\inst{1}and Jacques Zylberg\inst{1}}
\shortauthor{L. Bou\'e  \etal}

\institute{
  \inst{1} Department of Chemical Physics, The Weizmann
Institute of Science, Rehovot 76100, Israel\\
  \inst{2} School of Chemistry, The University of Sydney, NSW, 2006
Australia
}

\abstract{Developing a macroscopic theory of elasto-plasticity in amorphous solids calls for (i) identifying the relevant macro state-variables and (ii) discriminating the different time-scales which characterize these variables. In current theories it is assumed that the stress reaches its elasto-plastic steady state value on the same time-scale as the configurational variables (be they the configurational energy, configurational entropy or the effective temperature). By examining numerical simulations in two and three dimensions we show that this is generally not the case, the configurational degrees of freedom may reach the elasto-plastic steady state on the time scales which can be very different from the time scale of the stress. We provide a physical discussion to rationalize these findings.}

\begin{document}

\maketitle

\section{Introduction}
In contrast to fluids for which the Navier-Stokes equations provide an adequate description under a very wide range of conditions, we still do not have an accepted theory for amorphous solids that can describe their response to external loads from creep to steady elasto-plastic flows. The nonexistence of such an accepted theory is not for lack of trying. It has already been half a century since the pioneering work of Cohen and Turnbull \cite{59CT} and three decades since the early work of Spaepen and Argon \cite{77Spa,79Arg}, without an emerging theory that is accepted by one and all. In recent years the availability of more and more powerful computers has allowed crucial progress in understanding the intricacies of the subject \cite{04ML,04DA,06TLB,06ML,07BDLJ,09LP,09LC,09TTGB,09HKLP,09KLP}. In particular, numerical simulations can test all the basic assumptions that are made saliently or explicitly in various theories of the mechanical response of amorphous solids, allowing to weed out wrong assumptions or to validate fertile concepts. In this Letter we contribute to this path by focussing on the time-scales associated with relaxation to steady state of shear stress, potential energy and local structure in the context of elasto-plasticity.
\begin{figure}
\onefigure[scale = 0.55]{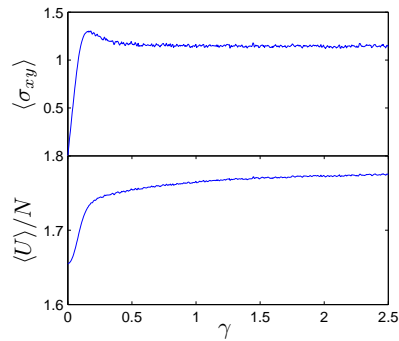}
\caption{Upper panel: Stress vs. strain curve in a 2-dimensional system (the hump model, defined below) of 6400 particles.
Data was averaged over 30 independent runs.
Lower panel: the energy approach to the steady state in the same system and the same conditions as in the upper panel. Note that the elastic energy reaches its steady-state value together with the stress, but then the configurational energy is
slow to attain the steady state.}
\label{stress-strain}
\end{figure}
Our central result is simply introduced. Consider a typical stress-strain curve (for numerical details see below) that is obtained in a 2-dimensional system of $N=6400$ particles with a strain rate of $\dot\gamma=10^{-4}$, cf. Fig.~\ref{stress-strain} upper panel. The stress reaches its steady state value, which is known as the flow-stress, at values of $\gamma$ of the order of $\gamma\approx 0.4$. On the other hand, the energy which is shown in the lower panel, reaches its
steady state value at much higher values of $\gamma$, values that in fact lie outside the range of this graph. Since $\dot\gamma$ is constant in this experiment, a value of $\gamma$ is also a time scale $\Delta t=\gamma/\dot \gamma$.
We impress on the reader two observations from Fig. \ref{stress-strain}: i) stress and energy approach their steady state values with quite different kinetics, and ii) following the rapid relaxation of the stress overshoot, the stress remains constant in time in spite of the fact that the energy (and, as we shall show, the microscopic structure) continue to evolve towards the steady state. These observations are not idiosyncrasies of a specific model. We shall demonstrate that this difference between stress and energy transients occurs generally across a range of models in 2D and 3D. We also made sure
that this result in not a consequence of shear-banding; the dynamics of our systems remains homogeneous at all strain values.

An appealing general approach to nonlinear rheology is to treat the behavior of the non-equilibrium state as if it was an equilibrium state but at an effective temperature. In this approximation, the role of the driving fields is reduced to establishing the appropriate effective temperature. An example of this approach is
the `Shear transformation Zone' theory of Langer and coworkers \cite{98FL,07MLC}, which is certainly one of the more attractive theories of elasto-plasticity that had been put forward in recent years. In this theory one asserts that elastic relaxation occurs due to the yielding of
shear transformation zones which are relatively rare and uncoupled groups of molecules whose
density is determined by an `effective temperature' which is a measure of the configurational
disorder. The effective temperature in this theory is the order parameter, and the time scale of
the stress reaching its steady state must be the same as the time scale for the effective temperature, or any other configurational variable, to reach its steady state. It is a fundamental
assumption of this theory (and in fact of any theory that assumes that an effective temperature is {\em the} order
parameter), that there cannot be a discrepancy between the time scales of the stress and of the configurational variables in reaching the steady state. The data shown in Fig.~\ref{stress-strain} is in clear contradiction to these assumption. Since the elastic contribution to the energy reaches its steady state together with the stress, we see that the {\em configurational} energy attains the steady state on a much slower time scale than the stress. This phenomenon is generic rather than specific to this model or another, as we show next.

\section{Models and numerical simulations}
To demonstrate the generality of the phenomena discussed we employ here
four different models of glass formation, all carefully studied before and all demonstrating the
usual properties of the glass transition and of the resulting amorphous solids. These models are
\begin{enumerate}
\item  Binary mixture with repulsive potential. Here the point particles interact via a purely repulsive
potential which diverges like $(\lambda_{ij}/r_{ij})^{10}$ for $r\to 0$ and which goes to zero continuously with two derivatives at a cutoff length $r=r_c$. There are three values of $\lambda_{ij}=1,1.18$ and 1.4 for the interaction between two `small'
particles, a `small' and a `large' particle and two `large' particles respectively. Details of the potential can be found in \cite{09LPCH,09LPa,09LPb}.
\item Repulsive potential with multi-dispersed length parameters.  Here again point particles interact via
the same potential as in the binary case, but the length parameter $2\lambda_{ij}=\lambda_i+\lambda_j$ where each
$\lambda_i$ is taken from a Gaussian distribution with $\frac{\langle \left(\lambda_i -
\langle \lambda \rangle \right)^2\rangle}{\langle \lambda \rangle^2}=15\%$. Details can be found in \cite{09LPCH,09LPa,09LPb}.
\item The Shintani-Tanaka model. This model is a glass-forming system whose constituents interact via an anisotropic potential depending on the angle of a unit vector carried by each particle. Details of the potential can be found in \cite{06ST,07ILLP,09LPR}.
\item The hump model. In this model the identical particles are interacting via a pair-wise potential that has
 a minimum, then a hump, and then it goes smoothly to zero at a finite distance $r_c$ with two derivatives.
 Details of the model can be found for example in \cite{09LPa,09BLPZ}
\end{enumerate}

\begin{figure}
\onefigure[scale = 0.60]{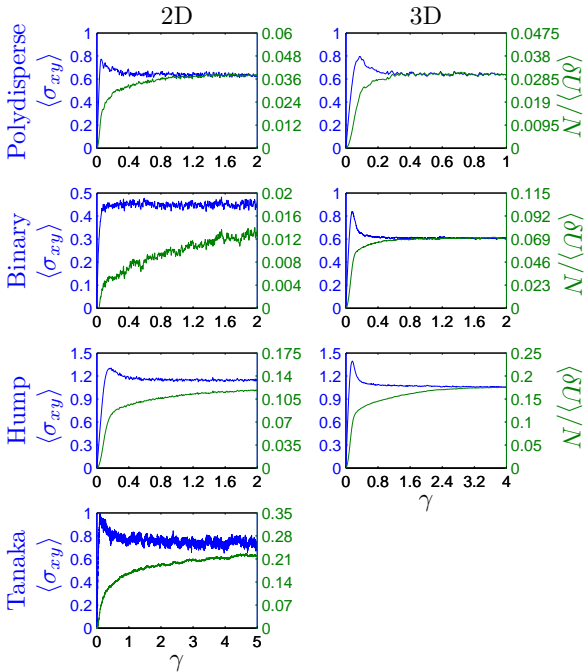}
\caption{Stress vs. strain curve and energy vs. strain as computed for models 2,1,3 and 4 respectively in 2D (left panels) and in 3D (right panels). The Tanaka model 4 is only defined in 2D. Note that the slow attainment of the configurational
energy to the steady state is very clear in 2D, but it exists in 3D as well, as seen for example in the hump model 3. The apparent identity of time scales in 3D between stress and energy in models 2 and 1 is accidental. All the simulations here and below employed the standard SLLOD algorithm.}
\label{2D-3D}
\end{figure}

One important difference between models 1 and 2 on the one hand and models 3 and 4 on the other is that the latter
 have a minimum in the potential where nearest-neighbor particles can reside for a long time. The former models
 are purely repulsive and can exchange nearest-neighbors on a vibrational time-scale. For all these models we have measured the stress and the energy approach to the steady state in two-dimensions (2D) and in three-dimensions (3D). Results are presented in Fig. \ref{2D-3D}.
Examining the results we conclude that (i) the observation that stress relaxes more quickly to its steady state value than does energy is generic, and (ii) this feature appears to be more pronounced in 2D than in 3D. To further underscore the generality of observation i), we note that Bulatov and Argon~\cite{94BA} reported similar results arising from a coarse grained model of plastic flow. We will argue next that this phenomenon can be ascribed in part to the stress being relatively
insensitive to the configurational degrees of freedom, whereas the energy, after reaching its elastic steady state,
is evolving further precisely on the time scale of the configurational degrees of freedom.

\section{The energy times scale and the configurational degrees of freedom}
\begin{figure}
\onefigure[scale = 0.60]{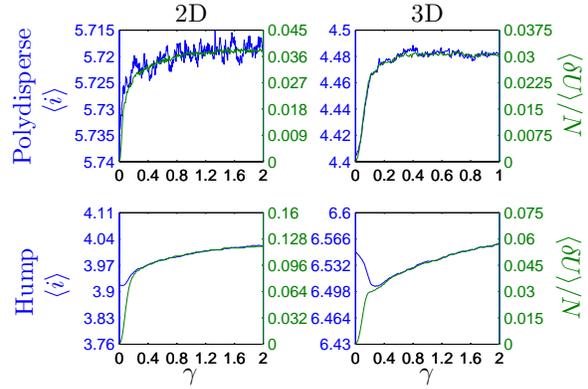}
\caption{Upper panels: Comparison between the attainment of steady state by the energy and by the mean number of neighbors. In 2D we employed a system with $N=20164$, in 3D with $N=16384$. Lower panels: the same comparison for the hump model, in 2D with a system of $N=6400$ and in 3D with $N=32768$.}
\label{neighbors}
\end{figure}

To demonstrate the relation between the time scales of the configurational degrees of freedom and the energy in attaining the steady state we use two approaches. In the first we simply recognize that changing the configurational degrees of freedom results in a temporal dependence of the mean number of nearest neighbors. In every model `nearest neighbors' may mean something slightly different; thus in the
multi-dispersed model a `neighbor' is any particle that resides within the interaction range, before the cut-off.
In the hump model a `neighbor' is counted as such if its distance is smaller than the position of the hump. Irrespective of these slight differences, one can see in Fig. \ref{neighbors} that the dynamics of the configurational energy follows
verbatim the dynamics of the average number of neighbors. In all cases the elastic energy reaches its asymptotic value rapidly, on the time scale of the stress, while the configurational energy follows the change in the average number of neighbors. Note the relatively higher fluctuations in the number of
neighbors in the multi-dispersed model 2 compared to the hump model 3; this follows from the definition of neighbors that can switch from one particle to the other on the time scale of a single vibration. For the hump model the switch is thermally activated and therefore much less readily made
at low temperatures. The other two models exhibit similar results, allowing us to conclude that for all models and all
dimensions the energy appears to follow the configurational change, and if its time scale appears in some simulations to be the same as that of the stress, this is accidental.

Along with the evolution of the average structure, we can also monitor the evolution of the non-affine character of the collective motions. To this end we denote the nearest neighbors of the $i$th particles
as $n.n.(i)$ and define the local non-affine deformation measure $q_i$ via \cite{98FL}:
\begin{equation}
q_i(\gamma,\gamma') \equiv \sum_{j \in \mbox{\tiny{n.n.($i$)}}}
\sum_\alpha \left(
r^{ij}_\alpha(\gamma)-
\Psi_{\alpha\nu}\Upsilon^{-1}_{\nu\beta}
r^{ij}_\beta(\gamma')\right)^2\ ,
\end{equation}
where $r^{ij}_\alpha \equiv r^{j}_\alpha - r^{i}_\alpha$, and
\begin{equation}
\begin{split}
\Psi_{\alpha\beta} = &\, \sum_{j \in \mbox{\tiny{n.n.($i$)}}}
r^{ij}_\alpha(\gamma)r^{ij}_\beta(\gamma')\ , \\
\Upsilon_{\alpha\beta} = &\, \sum_{j \in \mbox{\tiny{n.n.($i$)}}}
r^{ij}_\alpha(\gamma')r^{ij}_\beta(\gamma')\ .
\end{split}
\end{equation}
Clearly, $q_i \ge 0$ for any $\gamma,\gamma'$, and
is non-zero if the vicinity of particle $i$ undergoes
either a plastic deformation or non-affine elasticity.
The measure of non-affine deformation is then taken as $Q(\gamma)$:
\begin{equation}
Q(\gamma,\gamma') \equiv \sum_i q_i(\gamma,\gamma') \ .
\end{equation}

\begin{figure}
\onefigure[scale = 0.60]{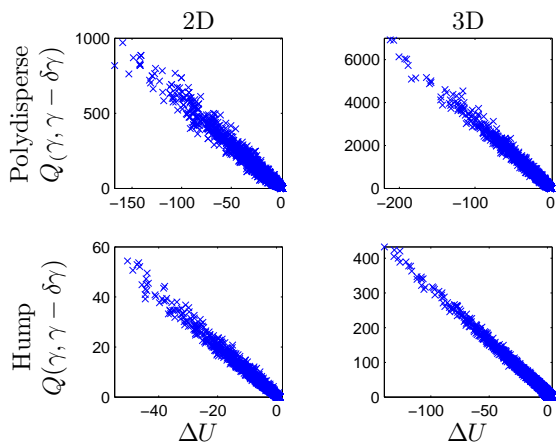}
\caption{Correlations between the energy drops and the increases in $Q(\gamma)$}
\label{QvsE}
\end{figure}

In order to demonstrate the correlation between the energy equilibration and the non-affine deformation processes, we
performed an athermal quasi-static
straining experiment, during which the energy change and the deformation measure $Q$ were computed. At each strain
state the energy change $\Delta U\equiv U(\gamma)-U(\gamma-\delta \gamma)$ and the measure of deformation $Q(\gamma, \gamma-\delta \gamma)$ are evaluated with respect to the previous recorded strained state in the
trajectory. In Fig. \ref{QvsE} we show the correlation plots between the
magnitude of the energy changes and the changes in the degree of mixing. We see the almost perfect correlations which
 are a general feature spanning across models and dimensions.

In summary, we showed that an explicit measure of the evolution of the configurations space of the shearing material (the average number of neighbors) and a measure of the character of collective particle movements (the non-affinity measure) are both strongly coupled to the evolution of the energy. We conclude that it is the energy, not the stress, that provides an accurate measure of the shearing system's progression towards its steady state.
The mechanism of the rapid attainment of the stress to its own steady state is still under
investigation and is not entirely understood. On the basis of the findings reported here, it appears that the
stress is much less coupled to the average configurational properties (such as an effective temperature) than is the energy. At this stage, we can only sketch a plausible physical account of these observations.  Imagine that the system has regions that are reluctant to relax, be they slow modes of the Hessian of the system or maybe even crystalline clusters that are very slow to disappear and mix \cite{08LP}. Such dynamical heterogeneities will
obviously affect the time scale of the attainment of configurational energy to the steady state. All parts of the system contribute to the energy. In this sense energy is a true reflection of the state of the entire system. Stress, in contrast, is determined by some subset of the system that may even live on a fractal. In any instantaneous configuration there exisit low energy (hard) and high energy (soft) regions. The stress is supported by the network of hard regions. The smaller the cross-sectional area of this net, the higher the stress for a given strain. This means that a relatively small increase in the amount of high energy (soft) regions can lead to a significant decrease in the cross-sectional area of the hard backbone and hence a large increase in stress. The evolution of stress is fast, in other words, because it reflects not only the time dependence of the amount of soft regions (i.e. the energy) but the more tenuous distribution of the remaining hard domains. This picture is supported at least qualitatively by the ability of the Bulatov and Argon model~\cite{94BA} to capture the different kinetics of stress and energy; they point to the importance of slip localization as an important contributing effect. Naturally, resolving fully the fundamental mechanisms for this complex dynamics will be the subject of future research. Finally, we emphasize that this paper is not intended as a global criticism of effective temperature. This concept may be valuable in characterizing the
elasto-plastic steady state itself \cite{09BHPRZ}. It is the relevance of the dynamics of the effective temperature to the
evolution of the stress that the results presented here challenge.

\acknowledgments

This work had been supported in part by the Israel Science Foundation and the German Israeli Foundation.

\end{document}